\documentclass[
  twoside,
  twocolumn,
  aps,
  10pt,
  prb,
  floatfix,
  showpacs,
  superscriptaddress,
  citeautoscript,
]{revtex4-1}

\usepackage{amsmath}
\usepackage{amssymb}
\usepackage{color}
\usepackage{graphicx}
\usepackage{tabularx}
\usepackage{subfigure}
\usepackage{dcolumn}
\usepackage{times}
\usepackage{glossaries}
\usepackage{units}

\graphicspath{{./figs/}}

\newcommand{\phys}{
  Chalmers University of Technology,
  Department of Physics,
  Gothenburg, Sweden
}

\newcommand{\sect}[1]{Sect.~\ref{#1}}
\newcommand{\fig}[1]{Fig.~\ref{#1}}
\newcommand{\eq}[1]{Eq.~(\ref{#1})}

\renewcommand{\vec}[1]{\ensuremath\boldsymbol{#1}}
\renewcommand{\epsilon}[0]{\varepsilon}

\newacronym{bcc}{BCC}{body centered cubic}
\newacronym{ce}{CE}{cluster expansion}
\newacronym{dft}{DFT}{density functional theory}
\newacronym{hcp}{HCP}{hexagonal closed-packed}
\newacronym{mc}{MC}{Monte Carlo}
\newacronym{md}{MD}{molecular dynamics}
\newacronym{sgc}{SGC}{semi-grand canonical}
\newacronym{vcsgc}{VCSGC}{variance constrained semi-grand canonical}
\newacronym{ardr}{ARDR}{automatic relevance detection regression}
\newacronym{vasp}{VASP}{Vienna ab initio simulation package}
\newacronym{lasso}{LASSO}{least absolute shrinkage and selection operator}
\newacronym{rfe}{RFE}{recursive feature elimination}
\newacronym{rmse}{RMSE}{root-mean-square error}
\newacronym{cv}{CV}{cross-validation}

\begin{document}

\preprint{}

\title{Structurally driven asymmetric miscibility in the phase diagram of W--Ti}

\author{Mattias {\AA}ngqvist}
\author{J. Magnus Rahm}
\author{Leili Gharaee}
\author{Paul Erhart}
\email{erhart@chalmers.se}
\affiliation{\phys}

\begin{abstract}
Phase diagrams for multi-component systems represent crucial information for understanding and designing materials but are very time consuming to assess experimentally.
Computational modeling plays an increasingly important role in this regard but has been largely focused on systems with matching lattice structures and/or stable boundary phases.
Here, using a combination of density functional theory calculations, alloy cluster expansions, free energy integration, and Monte Carlo simulations, we obtain the phase diagram of W--Ti, a system that features metastable boundary phases on both sides of the phase diagram.
We find that the mixing energy on the body-centered cubic (BCC) lattice is asymmetric and negative with a minimum of about \unit[$-120$]{meV/atom}, whereas for the hexagonal closed packed (HCP) lattice the mixing energy is positive.
By combining these data with a model for the vibrational free energy, we propose a revision of the W-rich end of the phase diagram with a much larger solubility of Ti in BCC-W than previous assessments.
Finally, by comparison with the W--V and W--Re system we demonstrate how strongly asymmetric phase diagrams can arise from a subtle energy balance of stable and metastable lattice structures.
\end{abstract}

\maketitle

\section{Introduction}

Metallic alloys play a crucial role in technology and are of continuing interest in basic research.
The most simple alloys comprise two components; these so-called binary systems can usually be categorized according to whether the interaction between the two constituents is repulsive (positive mixing energy) or attractive (negative mixing energy).
In the former case, examples for which include Cu--Ag \cite{Mas90, NajSroMa93, WilMisHam06} and Fe--Cu \cite{Mas90, HulDes71, JiaGenBor98, LopCarCar03}, one commonly observes a wide two-phase region (often referred to as the miscibility gap).
Attractive interactions, on the other hand, give rise to the formation of solid solutions, e.g., in Au--Ag \cite{Mas90, OzoWolZun98} or W--V \cite{Mas90, MuzNguKur11}, and the formation of intermetallic phases as in the case of Fe--Pt \cite{Mas90, LanBor96, FreSun01, KimKooLee06} or Ni--Al \cite{Mas90, AnsSunWil88, PurMis09}.

The interaction between the constituents is usually symmetric, in the sense that if $A$ dissolves in $B$, so does $B$ in $A$.
Exceptions from this behavior are rather uncommon; a prominent example is the Fe--Cr system \cite{CarCarLop06, BonPasMal09c}, for which the dissolution of Cr in Fe is energetically favorable whereas the opposite applies for Fe in Cr \cite{OlsAbrWal06, KlaDraFin06}.
As a result of this inversion in the mixing energy, the phase diagram is very asymmetric with a large solubility on the Fe-rich and a very small solubility on the Cr-rich side.
This behavior, which at first sight might be unexpected given a very small size mismatch and identical lattice structures, can be rationalized in terms of the magnetic structure \cite{Ack06, KlaDraFin06} with Fe and Cr preferring ferro and antiferromagnetic ordering, respectively.

Here, we show that strongly asymmetric phase diagrams can also be obtained in non-magnetic systems as a result of an asymmetry in lattice structures and their energetics.
In doing so, we also demonstrate the usage of the variance constrained semi-grand canonical Monte Carlo technique \cite{SadErh12, SadErhStu12} for extracting the complete free energy surfaces.
Specifically, we consider the W--Ti system.
Titanium exhibits a temperature driven transition from a low-temperature \gls{hcp} phase ($\alpha$-Ti) to a high-temperature \gls{bcc} phase ($\beta$-Ti) that is stabilized by vibrations \cite{PetHeiTra91}.
Tungsten is a refractory metal that maintains a \gls{bcc} structure up to the melting point.
Close-packed structures including \gls{hcp} are much higher in energy and can only be stabilized at very high pressures \cite{EinSadGri97, Ozo09}.

The W--Ti system is also of interest because its experimental assessment is aggravated by the high melting point of tungsten and the accompanying slow kinetics, which render the systematic exploration of the phase diagram, in particular the W-rich side, very cumbersome.
Since experimental data points for tungsten concentrations $\gtrsim\,30\%$ are only available down to 1473\,K \cite{RudWin68, Mur81}, the W--Ti system has been assessed using rather severe assumptions \cite{KauNes75, Mur81, LeeLee86, Jon96}.
It must be emphasized, however, that in spite of slow kinetics the low-temperature phase diagram of refractory alloy systems have a bearing, e.g., for the behavior under intense irradiation conditions such as in fusion reactors in so far as they determine the thermodynamic driving forces.
Tungsten alloys in particular are being considered for key components in fusion reactors that must sustain extreme mechanical and irradiation loads for prolonged periods of time \cite{ZinGho00, RieDudGon13, BecDom09}.

In the following, using a combination of \gls{dft} calculations, alloy \glspl{ce}, \gls{mc} simulations in the \gls{vcsgc} ensemble, and thermodynamic data for the pure elements \cite{Din91}, we provide a reassessment of the W--Ti phase diagram below the solidus line.
In doing so, we demonstrate (\emph{i}) that the solubility of Ti in W exceeds 20\%\ down to zero temperature in stark contrast to previous thermodynamic assessments \cite{KauNes75, LeeLee86, Jon96} while the inverse solubility is practically zero up to the \gls{hcp}-\gls{bcc} transition.
Furthermore, it is shown (\emph{ii}) that this asymmetry originates from a change of sign of the \gls{bcc}-\gls{hcp} free energy difference as a function of composition, which remains comparably small in magnitude.
In addition to providing a new perspective on the W--Ti system, the present results illustrate how strongly asymmetric phase diagrams can originate in non-magnetic systems, and illustrate a methodology that can handle multiple miscibility gaps in systems with several different lattice types.

The remainder of this paper is organized as follows.
The next section provides an overview of the computational techniques employed in this work.
The thermodynamic framework used to analyze the free energy landscapes and construct the phase diagram is outlined in \sect{sect:thermodynamic-modeling}.
The results of our computations are presented in \sect{sect:results}, after we discuss the implications for the W--Ti system in particular and put the present results in context in \sect{sect:discussion}.

\section{Computational methodology}
\label{sect:computational-methodology}

\subsection{Alloy cluster expansions}
\label{sect:cluster-expansions}

In the present work, we employ lattice Hamiltonians to represent the energy of the system as a function of composition and distribution of the elements.
These alloy \glspl{ce} can be written in the general form \cite{SanDucGra84}
\begin{align}
  \Delta E = \Delta E_{0}+ \sum_{\alpha}m_{\alpha} J_{\alpha} \overline{\Pi}_{\alpha} (\boldsymbol{\sigma}),
\end{align}
where $\Delta E$ denotes the mixing energy.
The summation runs over all symmetry inequivalent clusters $\alpha$ with multiplicity $m_{\alpha}$ and effective cluster interaction (ECI) $J_{\alpha}$.
The cluster correlations $\overline{\Pi}_{\alpha}$ are computed as symmetrized averages of products over the pseudospin vector $\boldsymbol{\sigma}$.
The latter represent the occupation of lattice sites by W ($\sigma=-1$) and Ti ($\sigma=+1$).

In the present work, we employed the \textsc{icet} package \cite{AngMunRah19} for the construction and sampling of \glspl{ce}.
We considered cluster spaces with up to 220 distinct clusters including clusters up to fifth order (quintuplets).
Inclusion of clusters of high order was necessary due to the very asymmetric shape of the mixing energy as a function of composition (see below).

For training of CEs, we systematically enumerated all structures with up to 12 atoms in the unit cell \cite{HarFor08, HarFor09}, which yields 10,846 and 5,777 structures for \gls{bcc} and \gls{hcp} lattices, respectively.
Based on this pool of structures, we relaxed and evaluated the energy of more than 1,700 \gls{bcc} and 900 \gls{hcp} structures using \gls{dft} as described in \sect{sect:density-functional-theory}.
Since both \gls{bcc}-Ti \cite{PetHeiTra91} and \gls{hcp}-W are structurally unstable, several configurations did not maintain their initial lattice structure.\footnote{We note that in the case of the related W--Re system it has been shown that the \gls{bcc} lattice remains dynamically stable up to approximately 70\%\ Re \cite{EkmPerGri00}.}
To exclude these structures, each configuration was mapped backed onto its respective ideal lattice structure.
\gls{bcc} structures were removed from the pool of structures if the displacement of any atom exceeded \unit[0.25]{\AA} or the tetragonal shear exceeded by 0.2\footnote{The tetragonal shear strain is defined as $\varepsilon_\mathrm{tet} = (\varepsilon_1+\varepsilon_2)/2\varepsilon_3$, where $\varepsilon_i$ are the eigenvalues of the strain tensor in Voigt notation ordered by magnitude $\varepsilon_1>\varepsilon_2>\varepsilon_3$.}, resulting in a total of 1,133 structures.
\gls{hcp} structures were removed if a structure had a negative mixing energy or a Ti concentration below 50\%, resulting in 105 structures.

We have previously shown that the \gls{ardr} optimization algorithm yields sparse solutions with low \gls{cv} scores, often outperforming both \gls{lasso} and \gls{rfe} approaches \cite{AngMunRah19}.
The $\lambda$-threshold parameter in \gls{ardr} controls the sparsity of the model with smaller values producing sparser solutions at the cost of higher validation-\gls{rmse} scores.
Sparser solutions are commonly both more transferable and also computationally less expensive to sample.
We therefore increased $\lambda$ until the \gls{rmse} score converged.
Here, \gls{cv} scores were estimated by the shuffle-and-split method with 50 splits using 90/10\%\ of the structures for training/validation.

\subsection{Monte Carlo simulations}
\label{sect:monte-carlo-simulations}

The final \glspl{ce} were sampled using Monte Carlo (MC) simulations.
In order to be able to construct the full free energy landscape of the crystalline phases (see \sect{sect:thermodynamic-modeling} below), we require the free energy for \gls{bcc} and \gls{hcp} phases separately, as a \emph{continuous} function of composition.
This prevents us from using the semi-grand canonical (SGC) ensemble.
While the latter does provide access to the first derivative of the free energy with respect to composition, it does not allow sampling multiphase regions, which as will be seen below are present for both \gls{bcc} and \gls{hcp} lattices.
To overcome this limitation we employ the \gls{vcsgc} ensemble.
It includes an additional term in the partition function that effectively imposes a constraint on the fluctuations of the concentration, which diverge in multi-phase regions.
The \gls{vcsgc}-\gls{mc} approach has been successfully employed previously to describe multiphase equilibria in, e.g., Fe--Cr \cite{SadErh12} and Fe--Cu alloys \cite{ErhSad13, ErhMarSad13}.
The \gls{vcsgc} ensemble is sampled by randomly selecting a site in the system, swapping its chemical identity, and accepting this trial move with probability \cite{SadErh12}
\begin{align}
  \mathcal{P} = \min\Big\{1,
    \exp\big[
      &-\beta \Delta E \\
      &- \kappa \Delta N_B \left( \phi + \Delta N_B / N + 2 N_B / N \right)
    \big]
  \Big\}.
  \nonumber
\end{align}
Here, $\Delta E$ is the energy change associated with the move, $\Delta N_B$ is the change in the number of particles of type $B$, $N$ is the total number of sites (atoms) in the simulation cell, and $\phi$ and $\kappa$ are the average and variance constraint parameters.
We employed $\kappa=200$ throughout; this choice provides a contiguous sampling of the concentration axis while maintaining a high acceptance rate and, in our experience, works almost universally for the systems that we have considered so far.
The average constraint parameter $\phi$ was varied in steps of 0.02 from $-2.2$ to $0.2$.
In the VCSGC ensemble the first derivative of the free energy is related to the (ensemble) average of the concentration $\left<c_B\right>=\left<N_B\right>/N$,
\begin{align}
  \beta \partial \Delta F/\partial c &= \kappa \left(\phi + 2 \left<c_B\right> \right),
  \label{eq:free-energy-vcsgc}
\end{align}
which allows one to obtain the free energy of mixing.

\gls{mc} simulations were carried out at temperatures between 300 and 1800\,K in 100\,K intervals using $5\times5\times5$ and $4\times4\times4$ supercells of the primitive unit cell for \gls{bcc} and \gls{hcp} structures, respectively.
At each value of $\phi$ the configuration was equilibrated for 10,000 trial steps, followed by 90,000 trial steps for gathering statistics.

\subsection{Density functional theory calculations}
\label{sect:density-functional-theory}

\begin{figure}
  \centering
  \includegraphics{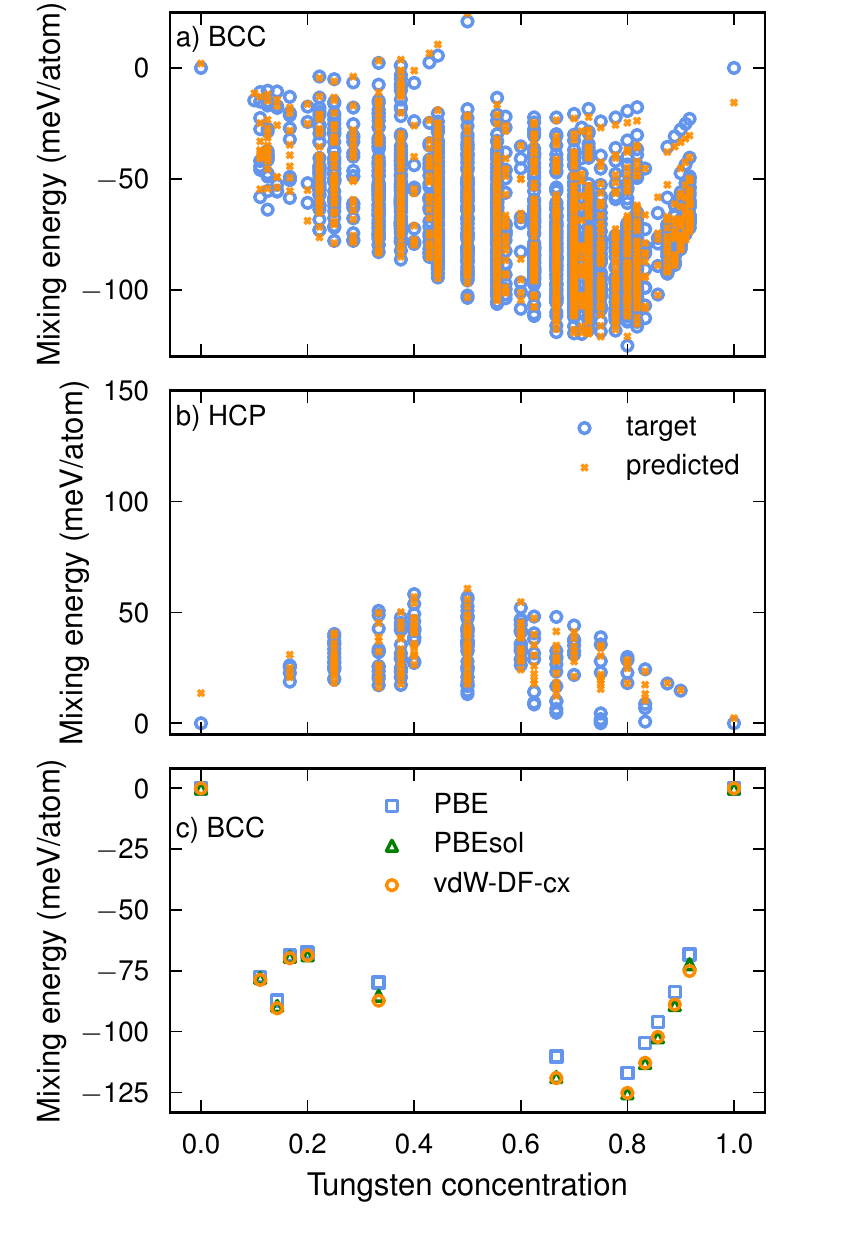}  
  \caption{
 Mixing energy for all (a) \gls{bcc} and (b) \gls{hcp} structures, for which \gls{dft} reference calculations were performed.
 (c) Comparison of the mixing energy computed using different exchange-correlation functionals for selected \gls{bcc} structures.
  }
  \label{fig:mixing-energy}
\end{figure}

\begin{figure}
  \centering
  \includegraphics{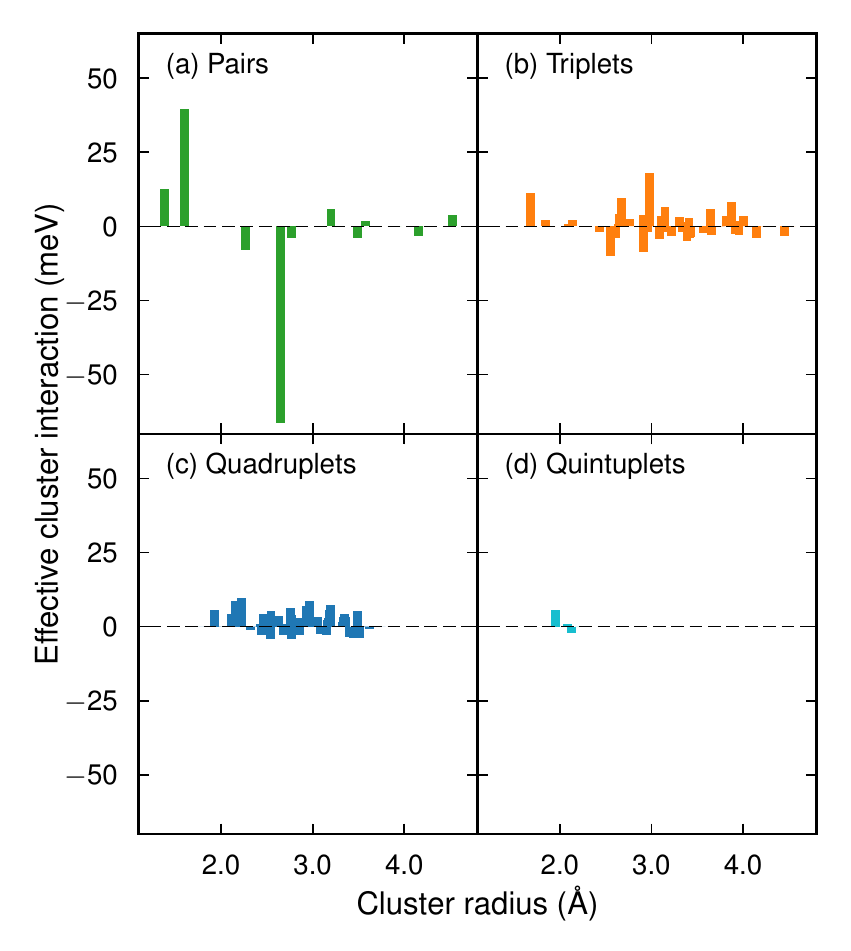}
  \caption{
    Effective cluster interactions (ECIs) for the cluster expansion for the \gls{bcc} lattice.}
  \label{fig:cluster-expansions}
\end{figure}

Density functional theory (DFT) calculations were carried out using the projector augmented wave method (PAW) \cite{Blo94, KreJou99} as implemented in the \gls{vasp} \cite{KreHaf93, KreFur96b}.
Electronic semi-core states (W-$5p$, Ti-$3p$) were treated as part of the valence and calculations employed a plane wave energy cutoff of 450\,eV.
Brillouin zone integrations were carried out using $\Gamma$-centered $\vec{k}$-point grids with an average spacing of 0.15/\AA, corresponding to a $19\times19\times19$ mesh relative to the primitive \gls{bcc} cell and a $18\times18\times9$ mesh relative to the primitive \gls{hcp} cell.
These computational settings are very tight and have been shown previously to yield well converged results \cite{GhaErh15}.

Using first-order Methfessel--Paxton smearing with a width of 0.2\,eV, both ionic positions and cell shapes were relaxed until residual forces were below 5\,meV/\AA\ and stresses below 0.5\,kbar.
The total energy of the final structures was subsequently computed without further relaxation using the tetrahedron method with Bl\"ochl corrections.

Exchange and correlation effects were described using the van der Waals density functional (vdW-DF) method that captures nonlocal correlations \cite{RydDioJac03, DioRydSch04} in combination with a consistent description of exchange (vdW-DF-cx) \cite{BerHyl14} as implemented in \gls{vasp} \cite{KliBowMic11, Bjo14}.
We have recently established that this functional provides a description of the thermophysical properties of non-magnetic transition metals that is at least on par with but usually exceeds other constraint-based functionals \cite{GhaErhHyl17}, notably PBE \cite{PerBurErn96} and PBEsol \cite{PerRuzCso08}.
To assess the effect of the exchange-correlation functional, we conducted supplementary calculations using the latter functionals for selected structures close to the \gls{bcc} convex hull.

\subsection{Thermodynamic methodology}
\label{sect:thermodynamic-modeling}

In order to construct the phase diagram, we consider the free energy landscape of the W--Ti system as a function of temperature $T$, W concentration $c$, and structure $\alpha$.
From the \gls{mc} simulations described above, we can extract the mixing free energy $\Delta G^\text{mix}_\alpha(c, T)$ for $\alpha=\text{BCC}$ or \gls{hcp}, which includes the contributions due to mixing energy $\Delta H^\text{mix}_\alpha(c,T)$ and configurational entropy $\Delta S^\text{mix}_\alpha(c, T)$.
Here, since ionic and cell relaxations are implicitly included in the ECIs, we have $\Delta G^\text{mix}_\alpha\approx \Delta F^\text{mix}_\alpha$ and from here on we will therefore refer only to $\Delta G$ and $\Delta H$.

To obtain the full Gibbs free energy, we must also take into account the vibrational contributions, which are not accounted for in the CE.
Since a full evaluation of the vibrational contribution as a function of composition is very demanding due to the importance of anharmonic contributions for Ti-rich \gls{bcc} and W-rich \gls{hcp} structures, we here approximate the vibrational contribution by a linear interpolation of the elemental free energies plus a correction for the strong anharmonicity of the metastable \gls{bcc}-Ti phase.
We thus write the total Gibbs free energy for phase $\alpha$ as
\begin{align} \label{eq:free-energy}
  G_\alpha(c, T)
  = c ~ \left[ G_\alpha^\text{W}(T) - G_\alpha^\text{W}(\unit[298.15]{K}) \right] \\
  + (1-c) ~ \left[ G_\alpha^\text{Ti}(T) - G_\alpha^\text{Ti}(\unit[298.15]{K}) \right] \nonumber \\
  + G_\alpha^\text{CE}(c) + \Delta H_\alpha^\text{anh}(c) \nonumber 
\end{align}
and for the sake of visualization the free energy of mixing as
\begin{align}
    \Delta G_{\alpha,\text{mix}}(c, T) = G_\alpha(c, T) \\
    - c ~ G_\text{BCC}(1, T) - (1-c) ~ G_\text{HCP} (0, T). \nonumber
\end{align}
For both $G_\alpha^\text{W}(T)$ and $G_\alpha^\text{Ti}(T)$, we resort to thermodynamic assessments available from the CALPHAD framework \cite{Din91}.
Equation~\eqref{eq:free-energy} contains an anharmonic correction term $\Delta H_\alpha^\text{anh}(c)$, which is motivated by an analysis of the lattice energy of \gls{bcc} structure that was carried out using \emph{ab-initio} molecular dynamics (MD) simulations with the \gls{vasp} package and the PBE exchange-correlation functional \cite{PerBurErn96}.
In these simulations, supercells comprising 54 atoms were sampled with XXX timesteps of \unit[1]{fs} using a Nos\'e--Hoover thermostat.

\section{Results}
\label{sect:results}

\subsection{Cluster expansions of mixing energies}

The \gls{dft} calculations reveal a negative mixing energy for \gls{bcc} structures [\fig{fig:mixing-energy}(a)].
The shape of the mixing energy is very asymmetric with several structures along the convex hull.
The lowest mixing energy is obtained for a structure at 80\%\ W.
This structure has space group 166 (R$\bar{3}$m) and contains 5 atoms in the primitive unit cell.\footnote{
  The lattice parameters of the ground state structure at 80\%\ W are $a=4.447\,\text{\AA}$, $c=13.708\,\text{\AA}$, when referred to the conventional (15-atom) unit cell.
The sole Ti atom is located at Wyckoff site $1a$ ($x,y,z=0,0,0$), while two tungsten atoms occupy two Wyckoff sites of type $2c$ ($x=y=z$) each with $x=0.20323$ and $x=0.40105$, respectively.
All parameters quoted here are from calculations using the vdW-DF-cx method and do not include zero-point motion.
}
For selected structures along the convex hull, we recalculated the mixing energies using the PBE and PBEsol functionals [\fig{fig:mixing-energy}(c)].
In particular between the PBEsol and vdW-DF-cx functionals the deviations are below one percent.

In the \gls{ce} formalism employed in the present work, reproducing the asymmetric shape of the mixing energy required including clusters up to fifth order and extending over rather long range [\fig{fig:cluster-expansions}(a)].
The final \gls{ce} to be used in the \gls{mc} simulations was obtained by training on the whole pool of structures with the \gls{ardr} optimization algorithm (see \sect{sect:cluster-expansions} for details).
The average root mean square error over the validation sets is 3.3\,meV/atom and the final \gls{ce} generally achieves very good agreement with the \gls{dft} reference data [\fig{fig:mixing-energy}(a)].
The accuracy and predictiveness of the \gls{ce} is also evident from the small errors of the mixing energy [\fig{fig:mixing-energy} (a)].

In contrast to the \gls{bcc} lattice, the mixing energy for the \gls{hcp} lattice is positive yet also asymmetric [\fig{fig:mixing-energy}(b)].
Here, the average root mean square error over the validation sets is 5.6\,meV/atom.

\subsection{Anharmonicity}

\begin{figure}
  \centering
  \includegraphics{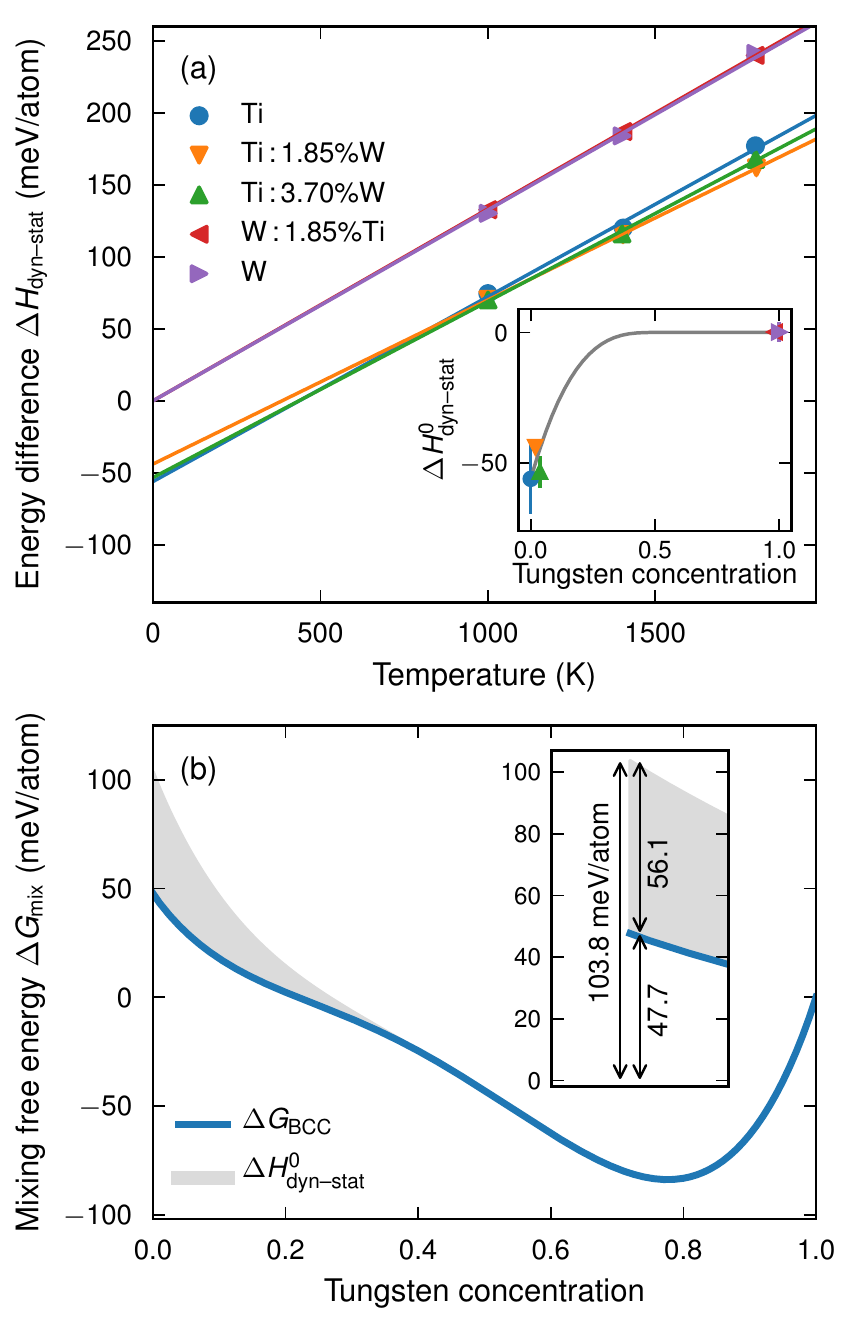}
  \caption{
    (a) Difference between the average potential energy at finite temperature and the static lattice energy in the \gls{bcc} phase for W--Ti at different concentrations as obtained from ab-initio \gls{md} simulations.
    The concentration dependence of this difference is in this work accounted for by \eq{eq:anharmonic-correction}, which is shown in the inset of (a).
    (b) Impact of the anharmonic correction (grey area) on the free energy of mixing (blue line) for the \gls{bcc} phase at \unit[298.15]{K}.
  }
  \label{fig:titanium-energies}
\end{figure}

In classical simulations the average potential energy of a system approaches its static value as the temperature goes to zero.
This behavior is observed for example for \gls{bcc}-W as simulated with ab-initio \gls{md} [\fig{fig:titanium-energies}(a)].
The average potential energy of the \gls{bcc}-Ti phase simulated in the same way does, however, not approach its static value but rather extrapolates to a value of $\Delta H_\text{dyn-stat, Ti}^0 = \unit[-56.1]{meV/atom}$.
This behavior originates from the metastable character of the \gls{bcc}-Ti phase and its mechanical instability at zero K, which implies a strongly anharmonic potential well.
By contrast, both \gls{bcc}-W and \gls{hcp}-Ti are stable at zero temperature and their dynamical behavior up to moderate temperatures can be comfortably described within the quasi-harmonic approximation \cite{GhaErhHyl17}.

The \gls{bcc}-CE is based on the static \gls{bcc}-Ti energy.
At finite temperatures, the free energy difference at 0\,K between the \gls{bcc}-Ti and \gls{hcp}-Ti phases is thus severely overestimated ($G^\text{CE}_\text{BCC-Ti} - G^\text{CE}_\text{HCP-Ti} = \unit[103.8]{meV/atom}$), unlike the CALPHAD assessment, which implicitly takes the correction into account ($G^\text{CALPHAD}_\text{BCC-Ti} - G^\text{CALPHAD}_\text{HCP-Ti} = \unit[47.3]{meV/atom}$ at the lowest available temperature).
By inclusion of the term $\Delta H_\text{dyn-stat, Ti}^0 = \unit[-56.1]{meV/atom}$, we obtain excellent agreement ($G^\text{CE}_\text{BCC-Ti} - G^\text{CE}_\text{HCP-Ti} + \Delta H_\text{dyn-stat}^0 = \unit[47.7]{meV/atom}$) with the CALPHAD assessment [\fig{fig:titanium-energies}(b)].

To achieve a consistent thermodynamic description we must account for the variation of $\Delta H_\text{dyn-stat}^0$ with composition, i.e., we must express $\Delta H_\alpha^\text{anh}(c)$ in \eq{eq:free-energy}.
Unfortunately, a comprehensive analysis of anharmonic behavior is already computationally demanding for the elemental phases \cite{HelSteAbr13, Ozo09, KadHonWal17}.
In the present case, we therefore make the basic assumption that $\Delta H_\alpha^\text{anh}(c)$ smoothly approaches $\Delta H_\text{dyn-stat, Ti}^0$ with decreasing Ti concentration.
We chose a simple cubic functional form [inset of \fig{fig:titanium-energies}(a)]
\begin{align}
  \Delta H_\text{BCC}^\text{anh} (c) = \Delta H_\text{dyn-stat, Ti}^0
  \begin{cases}
    (1 - c/c_s)^3 & c < c_s \\
    0 & c \geq c_s.
  \end{cases}
  \label{eq:anharmonic-correction}
\end{align}
Equation (\ref{eq:anharmonic-correction}) contains a single parameter $c_s$ that determines the concentration at which the correction is fully applied.
The thermodynamic analysis presented below was conducted using a value of $c_s=0.5$, motivated by qualitative agreement with ab-initio \gls{md} calculations of BCC-Ti with, respectively, 1.85\%\ and 3.70\%\ W [inset of \fig{fig:titanium-energies}(a)].

We note that a similar correction should in fact also be considered for \gls{hcp}-W vs \gls{bcc}-W.
As a result of the much larger energy difference between these structures, this effect is, however, insignificant for the phase diagram and has not been considered further (also see \sect{sect:discussion}).

\subsection{Free energy landscape}

\begin{figure}
  \centering
  \includegraphics{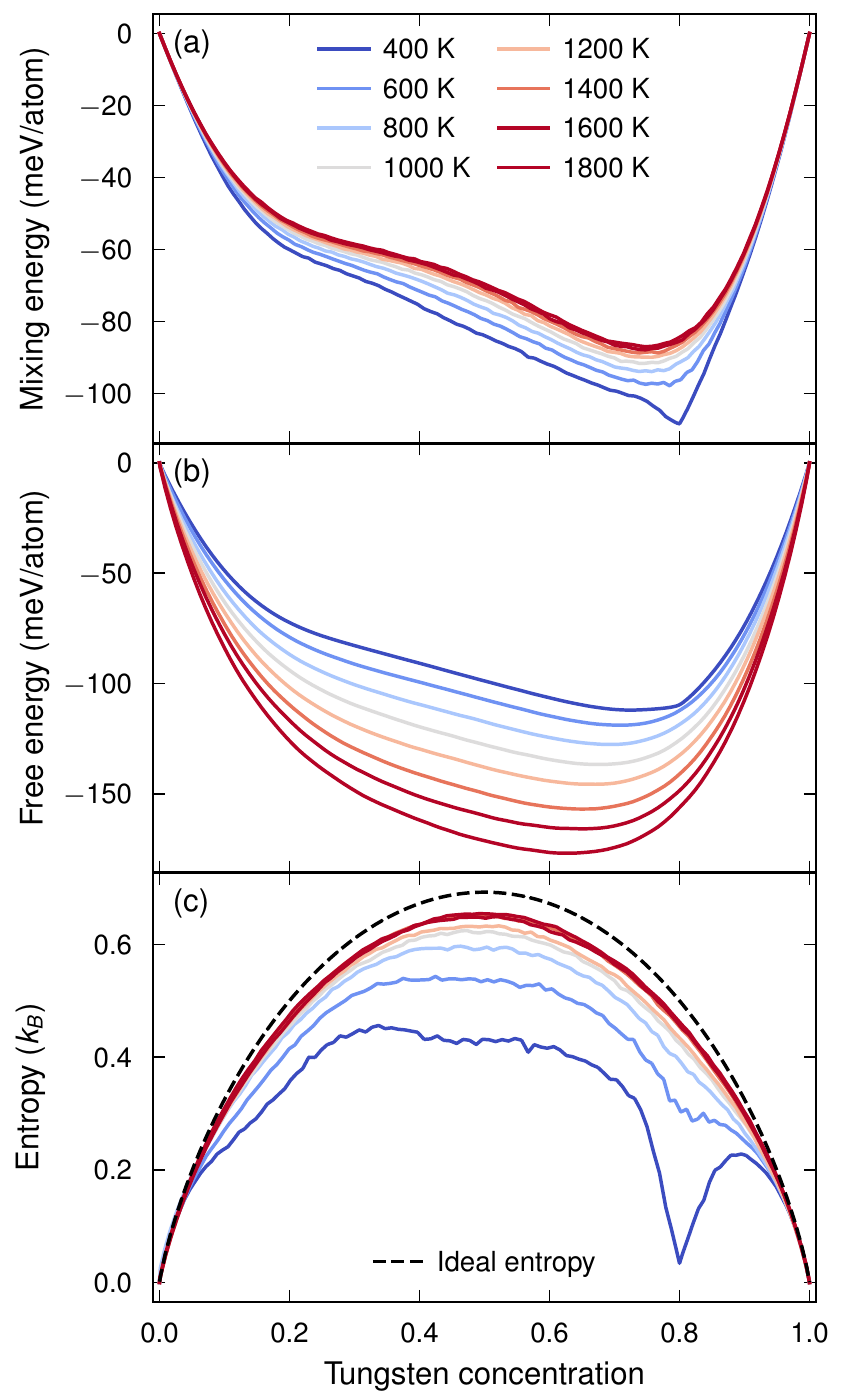}
  \caption{
    (a) Mixing energy, (b) free energy of mixing, and (c) mixing entropy as a function of composition from  \gls{vcsgc}-\gls{mc} simulations based on the \gls{ce} constructed for \gls{bcc} in this work.
  }
  \label{fig:free-energies}
\end{figure}

The \glspl{ce} for \gls{bcc} and \gls{hcp} lattices were sampled by \gls{mc} simulations as detailed in \sect{sect:monte-carlo-simulations}.
The (finite temperature) \gls{bcc} mixing energy maintains the asymmetric shape of the zero-temperature data [\fig{fig:free-energies}(a)].
It also clearly reveals the emergence of a particularly stable configuration at 80\%\ W, which corresponds to the ground state described above.

From the \gls{mc} simulations, we furthermore obtained the first derivative of the free energy with respect to concentration via \eq{eq:free-energy-vcsgc}, which was integrated using the trapezoidal rule to yield the mixing free energy [\fig{fig:free-energies}(b)].
Below the free energy of mixing will be used to construct the convex hull and the phase diagram.

By combining mixing free energy and energy, one can extract the entropy of mixing according to
\begin{align}
  \Delta S_\text{mix} = \left( \Delta H_\text{mix} - \Delta G_\text{mix} \right) / T.
\end{align}
At low temperatures the actual mixing entropy deviates strongly from that of an ideal solution and the ordered structure at 80\%\ W is clearly visible as a pronounced reduction in the mixing entropy [\fig{fig:free-energies}(c)].
At higher temperatures this feature is smoothed out and the mixing entropy becomes closer to that of an ideal solution.
Nonetheless, the very pronounced features in the entropy clearly demonstrate the importance of an accurate treatment of alloy thermodynamics beyond simple approximations.

In the case of the \gls{hcp} structures, mixing energies, free energies, and entropies were obtained in similar fashion as for the \gls{bcc} lattice.
Due to the positive mixing energy, these quantities are, however, much closer to those of an ideal solution and hence are not shown here explicitly.
Their determination is nonetheless important in order to obtain a consistent and comprehensive description of the thermodynamics of the relevant crystalline phases.

\subsection{Phase diagram}
\label{sect:phase-diagram}

\begin{figure}
  \centering
  \includegraphics{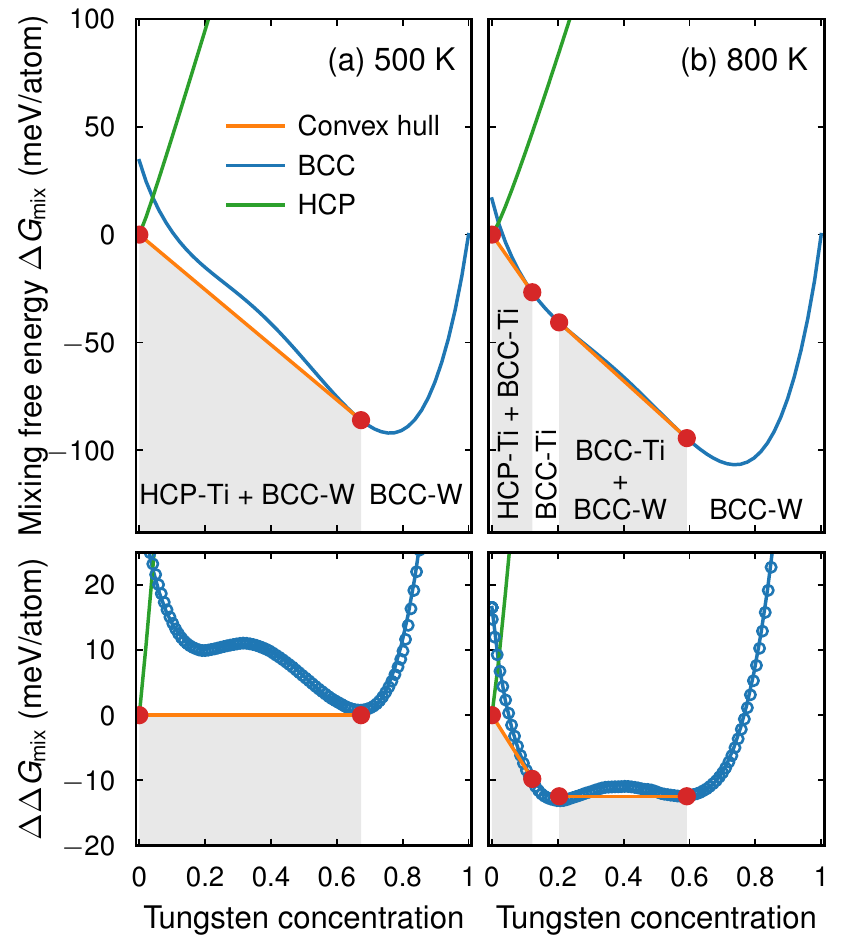}
  \caption{
    Mixing free energy as a function of composition at (a) 500 K and (b) 800 K.
The top row is based on the Redlich--Kister expansion of the free energy landscape obtained from \gls{vcsgc}-\gls{mc} simulations.
The data obtained from the latter are shown by the open circles in the bottom row in comparison with the expansion (solid blue line).
    For clarity of the visualization, in the bottom panel a linear term has been subtracted from the mixing energy, $\Delta\Delta G_\text{mix}(c) = \Delta G_\text{mix}(c) - m \cdot c$, where $m$ has been chosen to yield zero slope for the convex hull in the central two-phase regions.
The two-phase regions are highlighted in gray.
  }
  \label{fig:convex-hull}
\end{figure}

After having determined the free energies of both \gls{bcc} and \gls{hcp} phases, one can construct the full temperature, composition, and structure dependent free energy landscape, from which the phase diagram can be extracted.
To this end, the free energies for the different phases were combined as described in \sect{sect:thermodynamic-modeling}.
In accordance with experimental reality, our analysis yields three distinct stable crystalline phases, namely a Ti-rich \gls{hcp} phase (\gls{hcp}-Ti), a Ti-rich \gls{bcc} phase (\gls{bcc}-Ti), and a W-rich \gls{bcc} phase (\gls{bcc}-W).

At a temperature of 500\,K, the \gls{hcp}-Ti phase is in equilibrium with \gls{bcc}-W [\fig{fig:convex-hull}(a)] with the latter phase exhibiting a wide stability range with a solubility limit of 67\%\ W.
This substantially differs from the values $\gtrsim\,80\%$ predicted by CALPHAD assessments based on partial experimental data supplemented by approximations for the mixing energy \cite{LeeLee86, JinQiu93, Jon96}.
Similarly, at a temperature of 800\,K we obtain all three phases [\fig{fig:convex-hull}(b)].

In order to obtain a description of the phase diagram at all intermediate temperatures, we followed the common approach in alloy thermodynamics of representing the free energy of mixing in terms of a Redlich--Kister expansion
\begin{align}
  \Delta G(x, T) = x (1 - x) \sum_{p=0}^n L_p(T) (1 - 2 x)^p,
\end{align}
up to order $n=3$.
We then represented the temperature dependence of each of the eight (four per lattice) Redlich--Kister coefficients $L_p$ by third-order polynomials.
The interpolated mixing free energies are in very good agreement with our \gls{vcsgc}-\gls{mc} data [see points and blue lines in the bottom row of \fig{fig:convex-hull}], which allows us to extract the phase boundaries as a continuous function of temperature.

\begin{figure}
  \centering
  \includegraphics{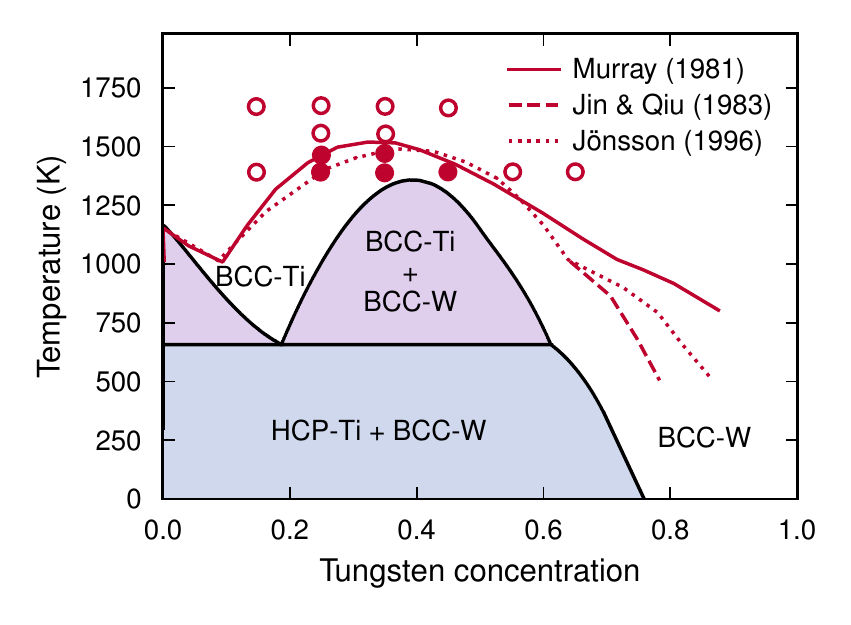}
  \caption{
    Phase diagram predicted based on the simulations in the present work in comparison with experimental data (symbols) from Ref.~\onlinecite{RudWin68} and thermodynamic assessments of the \gls{bcc}-W phase boundary (lines) from Refs.~\onlinecite{Mur81, JinQiu93, Jon96}.
Filled and empty squares indicate two-phase and single-phase regions, respectively.
  }
  \label{fig:phase-diagram}
\end{figure}

The thus obtained phase diagram [\fig{fig:phase-diagram}] shows the same phases and phase equilibria as the experimental one.
The upper temperature of the \gls{bcc}-Ti+\gls{bcc}-W two-phase region is predicted at \unit[1360]{K}, which is approximately \unit[160]{K} lower than experiment, and the position of the maximum is predicted at a W concentration of 39\%\ as opposed to approximately 30\%\ in the experiments.
The large high-temperature solubility results from a \emph{negative} mixing energy on the \gls{bcc} lattice, which also gives rise to \emph{finite} solubility limit of $>20\%$ for Ti in \gls{bcc}-W down to zero temperature.
The eutectoid on the Ti-rich side of the phase diagram is underestimated compared to experiment (657\,K vs 1013\,K), which comes with an overestimation of the eutectoid point (19\%\ vs 9\%).
The larger errors on the Ti-rich side are unsurprising given the difficulties associated with the strong anharmonicity of \gls{bcc}-Ti, which are only treated approximately in this work.
We note that we tested different values for $c_s$ parameter in \eq{eq:anharmonic-correction} and found the general shape of the phase diagram to be unaffected.
Larger (smaller) values of $c_s$ shift the \gls{bcc}-W solubility limit further to the W-rich side and tend to increase (decrease) the maximum of the \gls{bcc}-Ti+\gls{bcc}-W two phase region.

\section{Discussion and conclusions}
\label{sect:discussion}

\subsection{Relevance for the W--Ti system}

Above we have demonstrated that the solubility of Ti in \gls{bcc}-W at low temperatures is larger than previously predicted and remains finite as the temperature approaches zero.
This conclusion was reached by combining \gls{dft} calculations with effective lattice Hamiltonians, Monte Carlo simulations, and thermodynamic modeling.
By comparison, previous studies were based on experimental data, which is only available at temperatures above 1473\,K.
We note that while the low temperature region of the phase diagrams of refractory metals are exceedingly difficult to sample in equilibrium, they are nonetheless relevant as they determine the driving forces under extreme non-equilibrium situations such as encountered under ion irradiation.

\subsection{General implications}

While the present finding pertains to our understanding of the W--Ti phase diagram, it has more general implications for alloy thermodynamics.
As outlined in the introduction, binary phase diagrams of metals that exhibit strongly asymmetric solubility are relatively rare.
While as demonstrated in the case of the Fe--Cr system, they can arise from competing forms of magnetic order, here we show that this effect can also be observed in non-magnetic systems as a result of different lattice structures.

\begin{figure}
  \centering
  \includegraphics{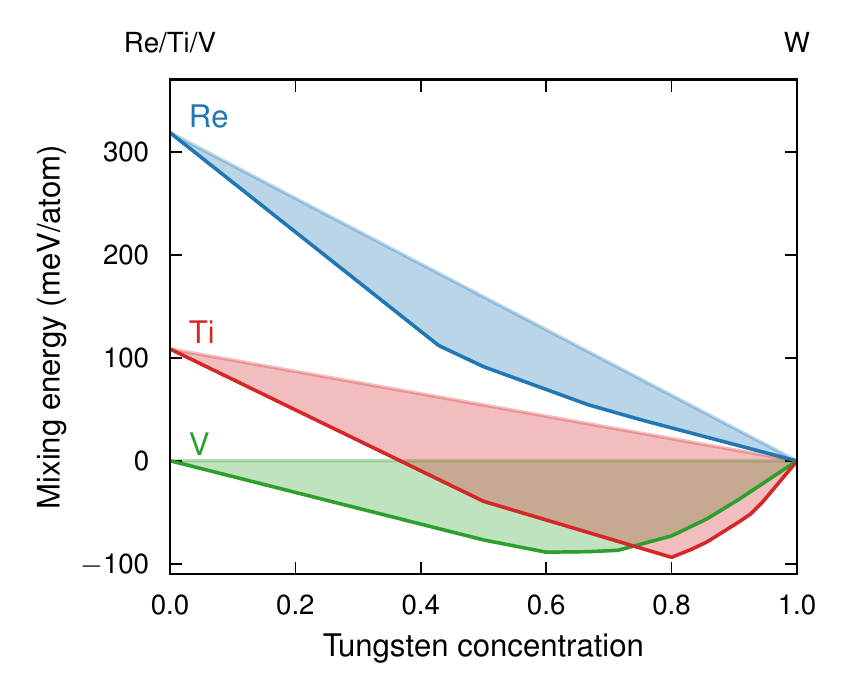}
  \caption{
    Schematic illustration of mixing energies for \gls{bcc} structures of W--V, W--Ti, and W--Re based on data from Ref.~\onlinecite{GhaMarErh16}.
In the case of the latter two alloys, the energy offset in the Ti/Re-rich limit represents the \gls{hcp}-\gls{bcc} energy difference.
The figure illustrate that the occurrence of a finite solubility at zero temperature in the case of W--Ti is the result of a combination of a negative mixing energy on the \gls{bcc} lattice and not-too-large \gls{hcp}-\gls{bcc} energy difference.
  }
  \label{fig:mixing-energy-schematic}
\end{figure}

To illustrate this effect consider the mixing energies of W--V (\gls{bcc}) and W--Re (\gls{bcc} and \gls{hcp}) in comparison with W--Ti (\fig{fig:mixing-energy-schematic}; data from Ref.~\onlinecite{GhaMarErh16}).
In the case of W--V both end members are \gls{bcc}, the mixing energy is negative and the phase diagram shows an extended miscibility range (and is symmetric).
In the case of W--Re, the calculations yield a negative mixing energy for \gls{bcc} and an almost vanishing mixing energy for the \gls{hcp} lattice, very similar to the case of W--Ti described above.
The (free) energy difference between \gls{hcp}-Re (the ground state) and \gls{bcc}-Re is, however, much larger than in the case of Ti.
As a result, the negative mixing energy of the \gls{bcc} lattice is shifted upward such that the solubility of Re in \gls{bcc}-W is rather small and approaches zero with vanishing temperature.

The comparison with W--V and W--Re demonstrates that the occurrence of an asymmetric phase diagram in W--Ti is the result of a negative mixing energy on the \gls{bcc} lattice in combination with a \gls{hcp}-\gls{bcc} energy difference that is not too large.
This reasoning implies that similar behavior can be expected in other systems that combine different lattice structures with elements that have metastable structures.
By extension, this should also apply to effective (``quasi'') phase diagrams between compounds.

\subsection{Computational modeling}

Using conventional sampling techniques based on either the semi-grand canonical or canonical ensembles, one cannot simultaneously sample multi-phase regions and derivatives of the free energy. \footnote{We note that the semi-grand canonical and canonical ensemble could be combined to carry out sampling and evolution of the system, respectively.
By alternating these techniques it would then in principle be possible to recover information about the first derivative of the free energy.
The resulting scheme is, however, costly as it would commonly require many more \gls{mc} cycles than the \gls{vcsgc}-\gls{mc} approach.} In the present work, however, we have tackled a system that combines different lattice structures with miscibility gaps.
This required the ability to obtain the free energy profile for the different phases as a continuous function of composition (and temperature).
Here, we were able to achieve this by using the \gls{vcsgc}-\gls{mc} approach, and have thereby demonstrated the power of this methodology to extract free energies and phase diagram information.
This suggests that in the future the \gls{vcsgc}-\gls{mc} method can be of great utility as the reach of \textit{ab initio} alloy thermodynamics widens to address more complex and demanding challenges.

\subsection{Short comings and outlook}

Arguably the biggest approximation adopted in the present work concerns the description of anharmonic contributions to the free energy.
As alluded to above, a direct assessment of the vibrational contributions to the free energy as a function of composition is computationally very demanding \cite{HelSteAbr13, Ozo09, KadHonWal17}.
While for vibrationally stable systems this can be achieved within the harmonic approximation \cite{WalCed02b, HuaGraZha16, ChiPanWal16}, a more comprehensive treatment of the anharmonicity in alloys is one of the remaining challenges in the development of computational alloy thermodynamics.

\begin{acknowledgments}
This work was supported the Swedish Research Council and the Knut and Alice Wallenberg foundation.
Com\-puter time allocations by the Swedish National Infrastructure for Computing at NSC (Link\"oping), C3SE (G\"oteborg), and PDC (Stockholm) are gratefully acknowledged.
\end{acknowledgments}

\end{document}